\documentstyle[aps,12pt,preprint]{revtex}
\def \to {\rightarrow}
\def \beq {\begin{equation}}
\def \eeq {\end{equation}}
\def \ba {\begin{eqnarray}}
\def \ea {\end{eqnarray}}
\def \jpsi {J/\psi}
\def \< {\left <}
\def \> {\right >}

\begin{document}
\baselineskip 20pt
\renewcommand{\thesection}{\Roman{section}}
~~
{\hfill PKU-TP-97-21}
\vskip 20mm
\begin{center}
{\Large \bf The Color-Octet intrinsic charm in $\eta^\prime$ and $B\to \eta^\prime X$ decays}
\end{center}

\vskip 10mm
\centerline{Feng Yuan}
\vskip 2mm
\centerline{\small {\it Department of Physics, Peking University, Beijing 100871, People's Republic
of China}} 
\vskip 4mm
\centerline{Kuang-Ta Chao}
\vskip 2mm
\centerline{\small {\it China Center of Advanced Science and Technology (World Laboratory), Beijing 100080,
People's Republic of China}}
\vskip 1mm
\centerline{\small {\it Department of Physics, Peking University, Beijing 100871, People's Republic of China}} %
\vskip 15mm

\begin{center}
{\bf\large Abstract}

\begin{minipage}{140mm}
\vskip 5mm
\par

Color-octet mechanism for the decay $B\to \eta^\prime X$ is proposed to explain
the large branching ratio of $Br(B\to \eta^\prime X)\sim 1\times 10^{-3}$ recently
announced by CLEO.
We argue that the inclusive $\eta^\prime$ production in $B$
decays may dominantly come from the Cabbibo favored $b\to (\bar c c)_8s$ process
where $\bar c c$ pair is in a color-octet configuration, and followed by the
nonperturbative transition $(\bar c c)_8\to \eta^\prime X$. The color-octet intrinsic
charm component in the higher Fock states of $\eta^\prime$ is crucial and is
induced by the strong coupling of $\eta^\prime$ to gluons via QCD axial anomaly.

\vskip 5mm
\noindent
PACS number(s): 13.25.Hw, 12.38.Lg, 14.40.Nd
\end{minipage}
\end{center}
\vfill\eject\pagestyle{plain}\setcounter{page}{1}

Recently CLEO has reported\cite{cleo} a very large branching ratio for the inclusive production
of $\eta\prime$ in the $B$ meson decay:
\beq
\label{exp}
Br(B\to \eta^\prime X;2.2GeV\le E_{\eta^\prime}\le 2.7GeV)=(7.5\pm 1.5\pm 1.1)\times 10^{-4}.
\eeq
If $\eta^\prime$ is regarded as a pure $SU(3)$ singlet meson made of $u$, $d$ and
$s$ quarks, the process $B\to \eta^\prime+X$ is Cabbibo suppressed by the
factor $V_{ub}$.
The conventional estimate of the branching ratio for this Cabbibo suppressed
process is of order of $10^{-6}$\cite{5251}, which is over 2 order of magnitude
smaller than the experimental result.
To explain the large branching ratio of $Br(B\to \eta^\prime+X)$,
some suggestions have been made\cite{5251,hou}. The authors of \cite{hou} give
arguments for the need of new physics to enhance the $b\to sg^*$
decays followed by $g^*\to \eta^\prime g$.
The latter step, $g^*\to \eta^\prime g$ is controlled by the effective
$\eta^\prime -g-g$ coupling via the QCD anomaly\cite{w-v}.
The authors of \cite{5251} suggest that it is the very strong coupling between $\eta^\prime$
and the color-singlet axial vector current of the charm quark that makes the dominant
contribution to the large $\eta^\prime$ production rate in the Cabbibo favored process
$b\to \bar c cs$.

In this note, we will discuss another possibility. Utilizing the unique feature of
$\eta^\prime$, {\it i.e.}, its strong coupling to gluons through QCD axial
anomaly, and the large coefficient for the color-octet $\bar cc$ current in the
$b\to \bar c cs$ process. We will argue 
that the production of $\eta^\prime$ in $B$ decays
may dominantly come from the Cabbibo favored $b\to (\bar c c)_8 s$ process where
$\bar c c$ pair is in a color-octet configuration followed by the color-octet
$\bar c c$ pair nonperturbatively evolving into $\eta^\prime$ via
$(\bar c c)_8 \to \eta^\prime +X$ due to the color-octet hidden charm component
in the higher Fock states of $\eta^\prime$.
The essential point of our argument is that due to the strong coupling of
gluons to the $\eta^\prime$ there should be an appreciable color-octet $(\bar c c)_8$
component mixed into the higher Fock states of $\eta^\prime$.

Before discussing this possibility, let us first recall the suggestion of \cite{5251},
where a large magnitude of the color-singlet ``intrinsic charm" of $\eta^\prime$ is
of critical importance. In our opinion, however, there could be two problems in this
approach: (1) the mixed decay constant $f_{\eta^\prime}^{(c)}$ suggested in
\cite{5251} may be too large; (2) the color-singlet matrix element
$|\langle \eta_c|\bar c\gamma_\mu \gamma_5 c|0\rangle|$ used in \cite{5251} may also
be too large.

To see these, we write the axial anomaly relation for the charm quark
\beq
\partial _\mu (\bar c\gamma_\mu\gamma_5 c)=2iM_c \bar c\gamma_5 c+\frac{\alpha_s}
        {4\pi}G_{\mu\nu}\tilde{G}_{\mu\nu},
\eeq
and define
\beq
\langle 0|\bar c\gamma_\mu\gamma_5 c|\eta^\prime (p)\rangle=if_{\eta^\prime}^{(c)} p_\mu.
\eeq
If the charm quark is infinitely heavy, the mixed decay constant $f_{\eta^\prime}^{(c)}$
should vanish. In the physical world, as argued in \cite{chao}, the nonvanishing
$f_{\eta^\prime}^{(c)}$ is due to the existence of the $\bar c c$ component
in the physical $\eta^\prime$ wave function, and then
\beq
f_{\eta^\prime}^{(c)}=O(\lambda_{\eta^\prime \eta_c} f_{\eta_c}),
\eeq
where $\lambda_{\eta^\prime \eta_c}$ is the mixing angle (in radian) between $\eta^\prime$
and $\eta_c$, and $f_{\eta_c}$ the decay constant of $\eta_c$. Based on the axial
anomaly and the gluonic matrix element of $\eta^\prime$, the mixing angle
$\lambda_{\eta^\prime \eta_c}$ has been estimated to be\cite{chao}
\beq
\lambda_{\eta^\prime \eta_c}\approx 1.2\times 10^{-2}.
\eeq
It reproduces the experimental value of $\Gamma(\jpsi\to \gamma \eta^\prime)$
via $\Gamma(\jpsi\to \gamma \eta_c)$ and $\eta^\prime-\eta_c$ mixing, and
is also consistent with most quark model calculations (see, {\it e.g.}, \cite{isgur}).
With $f_{\eta_c}\approx 400MeV$\cite{5251},
we find
\beq
f_{\eta^\prime}^{(c)}\ll f_\pi\approx f_{\eta^\prime}\approx 130MeV \cite{pdg},
\eeq
where $f_{\eta^\prime}$ is the flavor $SU(3)$-singlet decay constant of the
$\eta^\prime$. Including the mixing of $\eta^\prime$ with other $\bar c c$ states
like $\eta_c^\prime,~\eta_c^{\prime\prime},~\cdots$, will enhance
$f_{\eta^\prime}^{(c)}$.
As a safe estimate, we would expect
\beq
\lambda_{\eta^\prime-(\bar c c)}\approx (2-10)\%,
\eeq
and then $f_{\eta^\prime}^{(c)}\preceq 40 MeV$, where $\lambda_{\eta^\prime-(\bar c c)}$
is the total mixed amplitude of the $\bar c c$ component in $\eta^\prime$.
In any case, we think that $f_{\eta^\prime}^{(c)}=140MeV$, as suggested in\cite{5251},
is too large. Also, even with $f_{\eta^\prime}^{(c)}=40MeV$, we still have
$f_{\eta^\prime}^{(c)}\ll f_{\eta^\prime}$, and therefore in our estimate\cite{chao}
the relation
\beq
2iM_c\langle 0|\bar c\gamma_5 c|\eta^\prime\rangle \approx
        -\frac{\alpha_s}{4\pi}\langle 0|G_{\mu\nu}\tilde{G}_{\mu\nu}|\eta^\prime
        \rangle
\eeq
should be a fairly good and self-consistent approximation.

The second problem in \cite{5251} is the overestimate of the color-singlet
matrix element $|\langle 0 |\bar c\gamma_\mu\gamma_5 c|\eta_c\rangle |$, which
is related to $|\langle 0 |\bar c\gamma_\mu\gamma_5 c|\jpsi \rangle |$.
The latter is determined in \cite{5251} by exhausting the experimental value
of $\Gamma(B\to \jpsi+X)$. 
However, it is well known that at the lowest order the color-singlet part of
$b\to \bar c cs$ process only contributes one third of the decay width
$\Gamma(B\to \jpsi+X)$ measured by CLEO\cite{psi1,cleo2}.
The situation dose not get better even if the next-to-leading
order (NLO) corrections in $\alpha_s$ are included in the nonleptonic effective weak
Hamiltonian for $B$ decays\cite{nlo}.
If the NLO results are used, even with the large value of $f_{\eta^\prime}^{(c)}=140MeV$,
the estimated results of $\eta^\prime$ production 
in $B$ decays in \cite{5251} will become
\beq
\label{cs}
Br(b\to (\bar c c)_1+X;(\bar c c)_1\to \eta^\prime)=0.12\cdot 0.6 \cdot 0.9\times 10^{-3}
        =0.065\times 10^{-3},
\eeq
which is much smaller than the experimental result Eq.(\ref{exp}).

Because of the two problems mentioned above, the color-singlet intrinsic charm
mechanism suggested in \cite{5251} may encounter difficulties in explaining the
large $\eta^\prime$ production rate in $B$ decays. We therefore consider
another approach, {\it i.e.}, the color-octet intrinsic charm of the $\eta^\prime$.
Our suggestion is motivated by two physical aspects: (1) the color-octet production
for charmonium in $b\to \bar c cs$ process; (2) the appreciable color-octet
intrinsic charm component in the higher Fock states of $\eta^\prime$.

We now come to the production of color-octet components. It is known that
based on the NRQCD factorisation formalism\cite{nrqcd} the theoretical
investigations\cite{psi1,psi2} show that the color-octet
contributions might account for the main part of $\jpsi$ production in $B$
decays.
Encouraged by the success of the color-octet mechanism in the explanation of
$\jpsi$ production in $B$ decays, we conjecture that the color-octet contribution
may also explain the large branching ratio of $\eta^\prime$ production in $B$ decays.
To see the importance of the color-octet component in the Cabbibo favored
process $b\to \bar c c s$, we write explicitly the effective Hamiltonian of
nonleptonic $B$ decays\cite{psi1},
\ba
\label{hamilton}
\nonumber
H_{eff}&=&-\frac{G_F}{\sqrt{2}}V_{cb}V_{cs}^*\big ( \frac{2C_+-C_-}{3}
        \bar c\gamma_\mu (1-\gamma_5)c\bar s \gamma_\mu (1-\gamma_5)b\\
        &~&+
        (C_++C_-)\bar c\gamma_\mu(1-\gamma_5)T^a c \bar s\gamma^\mu(1-\gamma_5)
        T^a b\big ),
\ea
where $G_F$ is the Fermi constant and $V_{ij}$'s are KM matrix elements.
The coefficiets $C_+$ and $C_-$ are Wilson coefficients at the scale of $\mu=M_b$.
To leading order of $\alpha_s(M_b)$ and to all orders of $\alpha_s(M_b)ln(M_W/M_b)$,
they are
\ba
C_+(M_b)&\approx& [\alpha_s(M_b)/\alpha_s(M_W)]^{-6/23},\\
C_-(M_b)&\approx& [\alpha_s(M_b)/\alpha_s(M_W)]^{12/23}.
\ea
In this effective Hamiltonian Eq.(\ref{hamilton}), the first term is the
color-singlet term, while the second term is the color-octet term.
Numericaly, taking $\alpha_s(M_W)=0.116$ and $\alpha_s(M_b)=0.20$, the
coefficients in front of the two terms are then
\ba
C_1&=&\frac{2C_+-C_-}{3}=0.13,\\
C_8&=&C_++C_-=2.2.
\ea
The coefficient of the color-octet term is over $16$ times
larger than that of the color-singlet term.
So the color-octet contributions to the decay width of $B$ meson in the
$b\to \bar ccs$ process will have a factor of $\sim 280$ (squared the ratio
of $C_8/C_1$) enhancement compared to the color-singlet contributions.
This color-octet $b\to \bar ccs$ process may provide an explaination for
the $n_c$ and ${\cal B}_{s.l.}$ problems in $B$ decays\cite{co}.

For $\eta^\prime$ production, by including the intermediate $(\bar cc)_8$
contributions in the process $b\to \bar ccs\to \eta^\prime +X$, one might
explain the large branching ratio.
The color-octet contributions come from the following process
\beq
b\to (\bar c c)[{}^3S_1^{(8)}]s;{\rm ~~and~~ nonperturbatively}~~(\bar c c)[{}^3S_1^{(8)}]\to \eta^\prime +X.
\eeq
Here, the intermediate $\bar c c$ pair is in a color-octet ${}^3S_1$ configuration,
which comes from the hidden charm component in the higher Fock states
of $\eta^\prime$.
To realize the $(\bar c c)_8$ transition into $\eta^\prime$, one must introduce
the hidden charm component in the Fock states of $\eta^\prime$.
We write the Fock state expansion as
\beq
|\eta^\prime\rangle=\Psi_{\bar n n}|(\bar n n)_1\rangle+
        \Psi_{\bar nng}|(\bar n n)_8g\rangle+\Psi_{\bar ccg}|(\bar c c)[{}^3S_1^{(8)}]g\rangle+\cdots+
        \Psi_{\bar cc}|(\bar cc)_1\rangle+
        \cdots,
\eeq
where $|(\bar n n)_1\rangle=\frac{1}{\sqrt{3}}|(\bar uu+\bar dd+\bar ss)_1\rangle$
is the color-singlet and $SU(3)$ flavor-singlet state, which is the leading
part of the $\eta^\prime$ wave function (for convenience we neglect the
$\eta-\eta^\prime$ mixing).
The mixing of charm in $\eta^\prime$ can be understood by the transitions
$(\bar c c)_1\to two~ gluons\to (\bar n n)_1$ and
$(\bar c c)_8\to one~ gluon\to (\bar n n)_8$, as shown in Fig.1 $(a)$ and $(b)$
respectively.
A naive counting rule for the quark-gluon vertex might indicate
\beq
|\Psi_{\bar c cg}|>|\Psi_{\bar cc}|,
\eeq
because the former has fewer vertex and then is less OZI suppressed than the latter.
If this naive argument makes sense, the $\eta^\prime$ will have more color-octet
$\bar cc$ component than color-singlet $\bar cc$ component.
Then, because the short distance coefficient (Wilson coefficient) of the color-octet
contributions is over 16 times larger than that of the color-singlet contributions,
the former may dominate over the latter in the process $b\to \bar c c$ with
$\bar c c\to \eta^\prime X$.

We follow the calculations of $\jpsi$ production in $B$ decays\cite{psi1,psi2}
to estimate the production rate of $\eta^\prime$ in $B$ decays via $\bar c c$
transitions.
The color-singlet contribution has been estimated before in Eq.(\ref{cs}).
For the color-octet contribution, the calculation is simple, and the result is
\ba
\nonumber
\Gamma(B\to (\bar c c)[{}^3S_1^{(8)}]+X\to \eta^\prime +X)&=&
        \frac{\langle {\cal O}_8^{\eta^\prime}({}^3S_1[\bar c c])\rangle}
        {\langle {\cal O}_8^{\psi}({}^3S_1[\bar c c])\rangle}
        \frac{PS(\eta^\prime)}{PS(\psi)}
        \Gamma(B\to (\bar c c)[{}^3S_1^{(8)}]+X\to \jpsi +X) \\
\label{co}
     &=&\frac{\langle {\cal O}_8^{\eta^\prime}({}^3S_1[\bar c c])\rangle}
        {2M_c^2}(C_++C_-)^2(1+\frac{8M_c^2}{M_b^2})\hat\Gamma_0,
\ea
where
\beq
\hat\Gamma_0=|V_{cb}|^2(\frac{G_F^2}{144\pi})M_b^2M_c(1-\frac{4M_c^2}{M_b^2})
        (1-\frac{M_{\eta^\prime}}{M_b^2}).
\eeq
Here, $PS(\eta^\prime)$ and $PS(\psi)$ are the phase space factors of
processes $b\to \eta^\prime +X$ and $b\to \jpsi +X$ respectively.
In Eq.(\ref{co}), the notation $\langle {\cal O}_8^{\eta^\prime}({}^3S_1[\bar c c])\rangle$
is the long distance nonperturbative matrix element which represents the probability
of the $(\bar c c)[{}^3S_1^{(8)}]$ evolving into $\eta^\prime$.
To account for the experimental measurement $Br(B\to \eta^\prime X)\approx 1\times 10^{-3}$, the matrix element
$\langle {\cal O}_8^{\eta^\prime}({}^3S_1[\bar c c])\rangle$ must be taken
as $6.4\times 10^{-3}GeV^3$, which is three times smaller than
the color-octet matrix element
$\langle {\cal O}_8^{\psi}({}^3S_1[\bar c c])\rangle$ used in \cite{psi2}.

For the $\jpsi$, the Fock state expansion \cite{nrqcd}
\begin{eqnarray}
\nonumber
|J/\psi>&=&O(1)|c\bar{c}({}^3S_{1},\b{1})\rangle + O(v)|c\bar{c}({}^3P_{J'},\b{8})g\rangle \\
\nonumber
     &+&
    O(v^2)|c\bar{c}({}^3S_1,\b{8}~ or~ \b{1})gg\rangle+\cdots.
\nonumber
\end{eqnarray}
indicate the ${}^3S_1$ color-octet $\bar c c$ component in the higher Fock
states of $\jpsi$ is of order $O(v^2)\sim O(10^{-1})$, where
$v$ is the charm quark velocity in the charmonium. On the other hand,
${}^3S_1$ color-octet $\bar c c$ component in the higher Fock states of $\eta^\prime$
should be of order (see (17))
\beq
O (\Psi_{\bar ccg})> O (\Psi_{\bar cc})=(2-10)\times 10^{-2},
\eeq
as indicated by $\lambda_{\eta^\prime-(\bar cc)}=(2-10)\times 10^{-2}$ in Eq.(7).
This may imply that due to QCD axial anomaly the $\eta^\prime$ strongly
couples to gluons and then mixes substantially with $|(\bar cc)_8g\rangle$
and $|(\bar cc)_1\rangle$ Fock states via intermediate gluons.
Therefore, the color-octet $\bar cc$ matrix element for the $\eta^\prime$ is
not much smaller than the color-octet $\bar cc$ matrix element for the $\jpsi$.
This means that the obtained value for
$\langle {\cal O}_8^{\eta^\prime}({}^3S_1[\bar c c])\rangle$ by fitting the
$\eta^\prime$ data can be understood on the basis of QCD anomaly.

To conclude, in order to explain the large production rate of $\eta^\prime$
in $B$ decays within the Standard Model, the proposed color-singlet intrinsic
charm mechanism may encounter difficulties.
Therefore, we argue that the inclusive $\eta^\prime$ production in $B$
decays may dominantly come from the Cabbibo favored $b\to (\bar c c)_8s$ process
where $\bar c c$ pair is in a color-octet configuration, and followed by the
nonperturbative transition $(\bar c c)_8\to \eta^\prime X$. The color-octet intrinsic
charm component in the higher Fock states of $\eta^\prime$ is crucial and is
induced by the strong coupling of $\eta^\prime$ to gluons via QCD axial anomaly.
Further investigations will be made to examine this mechanism.

\vskip 1cm
\begin{center}
{\bf {\large {Acknowledgments}\ }}
\end{center}

This work was supported in part by the National Natural Science Foundation
of China, and the State Education Commission of China and the State
Commission of Science and Technology of China.

%\newpage

\vskip 20mm
\newpage
\centerline{\bf \large Figure Captions}
\vskip 1cm
\noindent
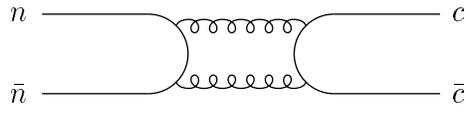
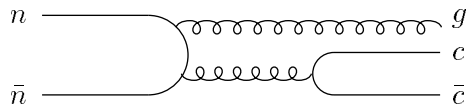
Fig.1. Hidden charm mixing in $\eta^\prime$ for the color-singlet $\bar c c$
pair $(a)$ and for the color-octet $\bar c c$ pair plus a gluon $(b)$.

\end{document}